\documentclass{article}

\begin{document}

\centerline{ A necessary and sufficient condition for genuinely entangled
 $n$-qubit states} \centerline{ with six non-zero coefficients} %
\centerline{Dafa Li}
\centerline{Department of
mathematical sciences, Tsinghua University, Beijing 100084 CHINA} %
\centerline{email: lidafa@tsinghua.edu.cn}

Abstract.  In [Science 340, 1205, 7 June (2013)], via polytopes Michael
Walter et al. proposed a sufficient condition detecting the genuinely
entangled pure states. In this paper, assume that a state with six non-zero
coefficients is not a trivially separable state. Then the state is separable
if and only if its six basis states consist of the three partially
complementary pairs\ and the corresponding coefficient matrix has
proportional rows. The contrapositive of this result reads that the state is
genuinely\ entangled if and only if its six basis states do not consist of
the three partially complementary pairs\ or though the six basis states
consist of the three partially complementary pairs, the corresponding
coefficient matrix does not have proportional rows. We propose four
corresponding coefficient 2 by 3 matrices and show that if the four
coefficient matrices don't have proportional rows, then the state is
genuinely entangled. It is trivial to know if two rows of a 2 by 3
coefficient matrix are proportional. The difference from the previous
articles is that the structure of the basis states is used to detect
entanglement in this paper. One can see that Osterloh and Siewert's states
of five and six qubits are genuinely entangled because two rows for any one
of the four corresponding coefficient 2 by 3 matrices are not proportional.
These states were distinguished as the maximal entangled states by the
complicated filters before.

Keywords: entanglement, separability, entangled states, separable states,
qubits.

\section{Introduction}

Quantum entanglement is a key physical resource in quantum information and
computation such as quantum secure communication, quantum cryptography,
quantum teleportation, quantum error correction code and so on. A pure state
of $n$ qubits\ is separable\ if it can be written as $|\varphi \rangle |\phi
\rangle $. Detecting genuine entanglement\ is a challenging task in quantum
information and computation \cite{Horodecki, Li-Pan}.

Many efforts have contributed to understanding the different ways of
entanglement \cite{Nielsen} and to deriving criteria for detection of
entanglement \cite{Peres}-\cite{Li-24}. Peres proposed the PPT criterion for
the separability of bipartite systems \cite{Peres}. It is known that the PPT
criterion for the separability is necessary and sufficient for 2 $\times $\
2 and 2 $\times $\ 3 systems \cite{Horodecki}. 3-tangle is sufficient for
detection of entanglement for three qubits. In \cite{Walter}, via polytopes
Michael Walter et al. gave the sufficient condition for the genuinely
entangled pure states.

The following techniques were used to detect genuine entanglement: linear
and nonlinear entanglement witnesseses \cite{Huber10}-\cite{Jungnitsch},
Bell inequalities \cite{Bancal, Zwerger}, and generalized concurrence \cite%
{Ming-Li-15, Chen, Gao}. Papers \cite{Shen, Jun-Li} discussed criteria to
detect and measure genuine bipartite and tripartite entanglement. We derived
necessary and sufficient conditions for genuine n-qubit entanglement \cite%
{Dli-prl, DLI-qip-19, Li-24} and reduced the detection of genuine
entanglement of n qubits to two qubits \cite{DLI-qip-21}.

Although it is hard to derive a general criterion for detection of
entanglement, it is possible to propose a necessary and sufficient condition
for genuinely entangled states with a few of non-zero coefficients. For
example, Osterloh and Siewert found the sets of the maximal entangled states
with a few of non-zero coefficients of four, five and six qubits by using
filters, respectively. In \cite{Li-24}, we gave a necessary and sufficient
condition for genuinely entangled states with four non-zero coefficients for
any $n$ qubits. In this paper, we propose a necessary and sufficient
condition for genuinely entangled $n$-qubit states with six non-zero
coefficients.

\section{Motivation}

It is possible to propose an approach to determine if a state with a few of
non-zero coefficients is genuinely entangled though it is hard to give a
general criteria to detect genuine entanglement. For example, the states
GHZ, W, Dicke, and cluster are genuinely entangled, they have a few of
non-zero coefficients. Osterloh and Siewert presented the entanglement
monotones \textquotedblleft filters\textquotedblright\ and found the
maximally entangled states of four, five, and six qubits\ with a few of
non-zero coefficients by using the filters \cite{Osterloh}. For example,
they gave the following two maximally entangled states with six non-zero
coefficients of five and six qubits.

$|\Psi _{6}\rangle =\frac{1}{2\sqrt{2}}(\sqrt{3}|11111\rangle +|10000\rangle
+|01000\rangle +|00100\rangle +|00010\rangle +|00001\rangle ),$

$|\Xi _{6}\rangle =\frac{1}{2\sqrt{2}}(\sqrt{3}|111111\rangle
+|110000\rangle +|00\rangle \otimes |W_{4}\rangle )$.

We proposed a necessary and sufficient condition to detect genuinely
entangled states with four non-zero coefficients for any $n$ qubits \cite%
{Li-24}. In this paper, we want to detect the genuinely entangled states of $%
n$ qubits with six non-zero coefficients. Let $|\psi \rangle _{1\cdots n}$
be a pure state of $n$ qubits\ and $m$\ be the number of non-zero
coefficients of $|\psi \rangle _{1\cdots n}$. When $m=6$,\ let $|\psi
\rangle _{1\cdots n}=\sum_{i=1}^{6}b_{i}|B_{i}\rangle _{1\cdots n}$, where $%
b_{i}\neq 0$, $i=1,..,6$ and $|B_{i}\rangle _{1\cdots n}$ are basis states,
where $B_{i}$ are binary numbers $\varepsilon _{1}^{(i)}\varepsilon
_{2}^{(i)}...\epsilon _{n}^{(i)}$, where $\varepsilon _{j}^{(i)}=0,1$, $%
i=1,\cdots ,6$, $j=1,\cdots ,n$, and
\begin{equation}
B_{1}<B_{2}<B_{3}<B_{4}<B_{5}<B_{6}.  \label{order}
\end{equation}

Let $1-\gamma =\gamma ^{\prime }$, where $\gamma =0,1$. Then, we can say
that $\gamma ^{\prime }$ and $\gamma $\ are complementary. When $|\psi
\rangle _{1\cdots n}$ $=|0\rangle _{i}|\varphi \rangle _{12\cdots n/i}$ ($%
|1\rangle _{i}|\phi \rangle _{12\cdots n/i}$), it means the ith qubit is 0
(1) in each basis state, $i=1,\cdots ,n$. Of course, the state is separable.
We call it a trivially separable state. It is easy to determine if a state
is trivially separable.

Let us consider the following state of four qubits
\begin{eqnarray}
|\eta \rangle _{1234} &=&\sum_{i=1}^{6}b_{i}|B_{i}\rangle _{1234} \\
&=&(|0000\rangle +|0001\rangle +|0101\rangle +|1010\rangle +|1011\rangle
+|1111\rangle )_{1234},
\end{eqnarray}%
where $b_{i}=1$, $i=1,\cdots ,6$.

One can see that $|\eta \rangle _{1234}$ can be written as
\begin{eqnarray}
|\eta \rangle _{1234} &=&(|00\rangle +|11\rangle )_{13}\otimes (|00\rangle
+|01\rangle +|11\rangle )_{24}  \label{eq-1} \\
&=&|00\rangle _{13}|00\rangle _{24}+|00\rangle _{13}|01\rangle
_{24}+|00\rangle _{13}|11\rangle _{24}+  \nonumber \\
&&|11\rangle _{13}|00\rangle _{24}+|11\rangle _{13}|01\rangle
_{24}+|11\rangle _{13}|11\rangle _{24}  \label{eq-3}
\end{eqnarray}

Then, we have the following three pairs: pair 1 $(00)_{13}(00)_{24}$ and $%
(11)_{13}(00)_{24}$; pair 2 $(00)_{13}(01)_{24}$ and $(11)_{13}(01)_{24}$;
pair 3 $(00)_{13}(11)_{24}$ and $(11)_{13}(11)_{24}$,\ which are referred to
as three partially complementary pairs

Generally, the following are three partially complementary pairs for $n$
qubits:
\begin{eqnarray}
&&(0_{1}\gamma _{2}\cdots \gamma _{k})_{p_{1}\cdots p_{k}}(0_{1}\sigma
_{2}\cdots \sigma _{s})_{q_{1}\cdots q_{s}},  \label{par-1} \\
&&(1_{1}\gamma _{2}^{\prime }\cdots \gamma _{k}^{\prime })_{p_{1}\cdots
p_{k}}(0_{1}\sigma _{2}\cdots \sigma _{s})_{q_{1}\cdots q_{s}}; \\
&&(0_{1}\gamma _{2}\cdots \gamma _{k})_{p_{1}\cdots p_{k}}(\tau _{1}\tau
_{2}\cdots \tau _{s})_{q_{1}\cdots q_{s}}, \\
&&(1_{1}\gamma _{2}^{\prime }\cdots \gamma _{k}^{\prime })_{p_{1}\cdots
p_{k}}(\tau _{1}\tau _{2}\cdots \tau _{s})_{q_{1}\cdots q_{s}}; \\
&&(0_{1}\gamma _{2}\cdots \gamma _{k})_{p_{1}\cdots p_{k}}(1_{1}\upsilon
_{2}\cdots \upsilon _{s})_{q_{1}\cdots q_{s}}, \\
&&(1_{1}\gamma _{2}^{\prime }\cdots \gamma _{k}^{\prime })_{p_{1}\cdots
p_{k}}(1_{1}\upsilon _{2}\cdots \upsilon _{s})_{q_{1}\cdots q_{s}},
\label{par-6}
\end{eqnarray}%
where $k+s=n$, $\sigma _{i}$, $\tau _{i}$, $\upsilon _{i}=0$ or $1$, $%
\{\sigma _{i}$, $\tau _{i}$, $\upsilon _{i}\}=\{1,0\}$, $i=1,\cdots ,s$.

Via the three pairs in Eqs. (\ref{par-1}-\ref{par-6}), we can define a state
of $n$ qubits as follows:

\begin{eqnarray}
|\chi \rangle _{1\cdots n} &=&(\alpha _{1}|0_{1}\gamma _{2}\cdots \gamma
_{k}\rangle +\alpha _{2}|1_{1}\gamma _{2}^{\prime }\cdots \gamma
_{k}^{\prime }\rangle )_{p_{1}\cdots p_{k}}\otimes  \nonumber \\
&&(\beta _{1}|0_{1}\sigma _{2}\cdots \sigma _{s}\rangle +\beta _{2}|\tau
_{1}\tau _{2}\cdots \tau _{s}\rangle +\beta _{3}|1_{1}\upsilon _{2}\cdots
\upsilon _{s}\rangle )_{q_{1}\cdots q_{s}}  \nonumber \\
&=&\alpha _{1}\beta _{1}|l_{1}\rangle _{1\cdots n}+\alpha _{1}\beta
_{2}|l_{2}\rangle _{1\cdots n}+\alpha _{1}\beta _{3}|l_{3}\rangle _{1\cdots
n}+  \nonumber \\
&&\alpha _{2}\beta _{1}|l_{4}\rangle _{1\cdots n}+\alpha _{2}\beta
_{2}|l_{5}\rangle _{1\cdots n}+\alpha _{2}\beta _{3}|l_{6}\rangle _{1\cdots
n}  \nonumber \\
&=&\sum_{i=1}^{6}\mu _{i}|l_{i}\rangle _{1\cdots n}.  \label{mod-1}
\end{eqnarray}

We call the following matrix the corresponding coefficient one, which is
denoted by $\Delta $. It is trivial to see that $\Delta $ has proportional
rows.
\begin{equation}
\Delta =\left(
\begin{array}{ccc}
\mu _{1} & \mu _{2} & \mu _{3} \\
\mu _{4} & \mu _{5} & \mu _{6}%
\end{array}%
\right) =\left(
\begin{array}{ccc}
\alpha _{1}\beta _{1} & \alpha _{1}\beta _{2} & \alpha _{1}\beta _{3} \\
\alpha _{2}\beta _{1} & \alpha _{2}\beta _{2} & \alpha _{2}\beta _{3}%
\end{array}%
\right)  \label{mod-2}
\end{equation}

It is clear that $|\chi \rangle _{1\cdots n}$ is separable but not trivially
separable because $\{\sigma _{i}$, $\tau _{i}$, $\upsilon _{i}\}=\{1,0\}$, $%
i=1,\cdots ,s$.

We will show that if the state $|\psi \rangle _{1\cdots
n}=\sum_{i=1}^{6}b_{i}|B_{i}\rangle _{1\cdots n}$ of $n$ qubits\ is a
separable but not trivial separable, then $|\psi \rangle _{1\cdots n}$ can
be rewritten as $|\chi \rangle _{1\cdots n}$. Assume that $|\chi \rangle
_{1\cdots n}=|\psi \rangle _{1\cdots n}$, one can see if $l_{i}=B_{j}$, then
$\mu _{i}=b_{j}$. Via the six coefficients $b_{1}$, $\cdots $, $b_{6}$, we
can make $6!(=720)$ $2$ by $3$ matrices with the entries $b_{i}$. We will
show that the corresponding coefficient matrices $\Delta $\ are of the
following four forms:
\begin{eqnarray}
&&\left(
\begin{array}{ccc}
b_{1} & b_{2} & b_{5} \\
b_{3} & b_{4} & b_{6}%
\end{array}%
\right)  \label{mt-1} \\
&&\left(
\begin{array}{ccc}
b_{1} & b_{3} & b_{5} \\
b_{2} & b_{4} & b_{6}%
\end{array}%
\right)  \label{mt-2} \\
&&\left(
\begin{array}{ccc}
b_{1} & b_{2} & b_{3} \\
b_{4} & b_{5} & b_{6}%
\end{array}%
\right)  \label{mt-3} \\
&&\left(
\begin{array}{ccc}
b_{1} & b_{3} & b_{4} \\
b_{2} & b_{5} & b_{6}%
\end{array}%
\right)  \label{mt-5}
\end{eqnarray}%
where for each of Eqs. (\ref{mt-1}, \ref{mt-2}, \ref{mt-3}, \ref{mt-5}), its
two rows are proportional. One can know that two rows are proportional if
and only if two columns are proportional.

\section{A necessary condition for separability of states of $n$ qubits with
six non-zero coefficients}

Lemma 1 (necessity). Assume that an $n$-qubit state $|\psi \rangle _{1\cdots
n}$ with $m=6$\ is not a trivially separable state. If $|\psi \rangle
_{1\cdots n}$ is separable, then its basis states consist of three partially
complementary pairs in Eqs. (\ref{par-1}-\ref{par-6}) and the corresponding
coefficient matrix has proportional rows.

Proof. The state can be written as $|\psi \rangle _{1\cdots
n}=\sum_{i=1}^{6}b_{i}|B_{i}\rangle _{1\cdots n}$. One can see that $|\psi
\rangle _{1\cdots n}$ must be a product state of a state with two non-zero
coefficients\ of qubits $p_{1},\cdots ,p_{k}$ and a state with three
non-zero coefficients of qubits $q_{1},\cdots ,q_{s}$, where $k+s=n$. Since $%
|\psi \rangle _{1\cdots n}$ is not a trivially separable state, there are
two cases to write $|\psi \rangle _{1\cdots n}$ below.

Case 1.

\begin{eqnarray}
|\psi \rangle _{1\cdots n} &=&(\alpha _{1}|0_{1}\gamma _{2}\cdots \gamma
_{k}\rangle +\alpha _{2}|1_{1}\gamma _{2}^{\prime }\cdots \gamma
_{k}^{\prime }\rangle )_{p_{1}\cdots p_{k}}\otimes  \nonumber \\
&&(\beta _{1}|0_{1}\sigma _{2}\cdots \sigma _{s}\rangle +\beta
_{2}|1_{1}\tau _{2}\cdots \tau _{s}\rangle +\beta _{3}|1_{1}\upsilon
_{2}\cdots \upsilon _{s}\rangle )_{q_{1}\cdots q_{s}},  \label{form-1}
\end{eqnarray}%
where $\tau _{2}\cdots \tau _{s}<\upsilon _{2}\cdots \upsilon _{s}$ without
loss of generality.

Case 2.
\begin{eqnarray}
|\psi \rangle _{1\cdots n} &=&(\alpha _{1}|0_{1}\gamma _{2}\cdots \gamma
_{k}\rangle +\alpha _{2}|1_{1}\gamma _{2}^{\prime }\cdots \gamma
_{k}^{\prime }\rangle )_{p_{1}\cdots p_{k}}\otimes  \nonumber \\
&&(\beta _{1}|0_{1}\sigma _{2}\cdots \sigma _{s}\rangle +\beta
_{2}|0_{1}\tau _{2}\cdots \tau _{s}\rangle +\beta _{3}|1_{1}\upsilon
_{2}\cdots \upsilon _{s}\rangle )_{q_{1}\cdots q_{s}},  \label{form-2}
\end{eqnarray}%
where $\sigma _{2}\cdots \sigma _{s}<\tau _{2}\cdots \tau _{s}$ without loss
of generality.

For Cases 1 and 2, $\{\sigma _{i},\tau _{i},\upsilon _{i}\}=\{1,0\}$, $%
i=1,...,s$, because $|\psi \rangle _{1\cdots n}$ is not a trivially
separable state.

For Case 1, Eq. (\ref{form-1}) can be rewritten as
\begin{eqnarray}
|\psi \rangle _{1\cdots n} &=&\alpha _{1}\beta _{1}|l_{1}\rangle _{1\cdots
n}+\alpha _{1}\beta _{2}|l_{2}\rangle _{1\cdots n}+\alpha _{1}\beta
_{3}|l_{3}\rangle _{1\cdots n}+  \nonumber \\
&&\alpha _{2}\beta _{1}|l_{4}\rangle _{1\cdots n}+\alpha _{2}\beta
_{2}|l_{5}\rangle _{1\cdots n}+\alpha _{2}\beta _{3}|l_{6}\rangle _{1\cdots
n}  \label{cf-1-0-} \\
&=&\sum_{i=1}^{6}\mu _{i}|l_{i}\rangle _{1\cdots n},  \label{cf-1-0}
\end{eqnarray}%
where
\begin{eqnarray}
|l_{1}\rangle _{1\cdots n} &=&|0_{1}\gamma _{2}\cdots \gamma _{k}\rangle
_{p_{1}\cdots p_{k}}|0_{1}\sigma _{2}\cdots \sigma _{s}\rangle _{q_{1}\cdots
q_{s}}, \\
|l_{2}\rangle _{1\cdots n} &=&|0_{1}\gamma _{2}\cdots \gamma _{k}\rangle
_{p_{1}\cdots p_{k}}|1_{1}\tau _{2}\cdots \tau _{s}\rangle _{q_{1}\cdots
q_{s}}, \\
|l_{3}\rangle _{1\cdots n} &=&|0_{1}\gamma _{2}\cdots \gamma _{k}\rangle
_{p_{1}\cdots p_{k}}|1_{1}\upsilon _{2}\cdots \upsilon _{s}\rangle
)_{q_{1}\cdots q_{s}}, \\
|l_{4}\rangle _{1\cdots n} &=&|1_{1}\gamma _{2}^{\prime }\cdots \gamma
_{k}^{\prime }\rangle )_{p_{1}\cdots p_{k}}|0_{1}\sigma _{2}\cdots \sigma
_{s}\rangle _{q_{1}\cdots q_{s}}, \\
|l_{5}\rangle _{1\cdots n} &=&|1_{1}\gamma _{2}^{\prime }\cdots \gamma
_{k}^{\prime }\rangle )_{p_{1}\cdots p_{k}}|1_{1}\tau _{2}\cdots \tau
_{s}\rangle _{q_{1}\cdots q_{s}}, \\
|l_{6}\rangle _{1\cdots n} &=&|1_{1}\gamma _{2}^{\prime }\cdots \gamma
_{k}^{\prime }\rangle )_{p_{1}\cdots p_{k}}|1_{1}\upsilon _{2}\cdots
\upsilon _{s}\rangle )_{q_{1}\cdots q_{s}}\rangle .
\end{eqnarray}

Clearly, $l_{1}$ and $l_{4}$, $l_{2}$ and $l_{5}$, and $l_{3}$ and $l_{6}$
are the desired three partially complementary pairs. Note that $\tau
_{2}\cdots \tau _{s}<\upsilon _{2}\cdots \upsilon _{s}$. Then, one can check
that $l_{2}<l_{3}$, $l_{4}<l_{5}$, and $l_{2}<l_{5}$.

We show that $l_{2}<l_{4}<l_{3}$ does not hold as follows. The following is
our argument. When $p_{1}$ is the qubit 1, $l_{2}<l_{4}$ but $l_{4}>l_{3}$.
When $q_{1}$ is the qubit 1, $l_{4}<l_{3}$ but $l_{2}>l_{4}$. Therefore, $%
l_{2}<l_{4}<l_{3}$ does not hold. Specially, $%
l_{1}<l_{2}<l_{4}<l_{3}<l_{5}<l_{6}$ does not hold.

We next show that it is not possible to make $l_{2}<l_{4}$ and $l_{5}<l_{3}$
together as follows. When, $p_{1}$ is qubit 1, $l_{2}<l_{4}$ but $%
l_{5}>l_{3} $. When $q_{1}$ is qubit 1, $l_{2}>l_{4}$. Therefore, it is not
possible for $l_{2}<l_{4}$ and $l_{5}<l_{3}$ to both hold. Specially, $%
l_{1}<l_{2}<l_{4}<l_{5}<l_{3}<l_{6}$ does not hold.

Therefore, $l_{i},i=1,..,6$, must satisfy one of the following:

\begin{eqnarray}
l_{1} &<&l_{4}<l_{2}<l_{5}<l_{3}<l_{6}  \label{rel-2} \\
l_{1} &<&l_{2}<l_{3}<l_{4}<l_{5}<l_{6}  \label{rel-3} \\
l_{1} &<&l_{4}<l_{2}<l_{3}<l_{5}<l_{6}  \label{rel-5}
\end{eqnarray}

From Eqs (\ref{cf-1-0}, \ref{rel-2}, \ref{order}), it is clear that $%
|l_{1}\rangle =|B_{1}\rangle $, $|l_{4}\rangle =|B_{2}\rangle $, $%
|l_{2}\rangle =|B_{3}\rangle $, $|l_{5}\rangle =|B_{4}\rangle $, $%
|l_{3}\rangle =|B_{5}\rangle $, and $|l_{6}\rangle =$ $|B_{6}\rangle $.
Thus, obtain the following corresponding coefficient matrix

\begin{eqnarray}
\left(
\begin{array}{ccc}
\alpha _{1}\beta _{1} & \alpha _{1}\beta _{2} & \alpha _{1}\beta _{3} \\
\alpha _{2}\beta _{1} & \alpha _{2}\beta _{2} & \alpha _{2}\beta _{3}%
\end{array}%
\right) &=&\left(
\begin{array}{ccc}
\mu _{1} & \mu _{2} & \mu _{3} \\
\mu _{4} & \mu _{5} & \mu _{6}%
\end{array}%
\right)  \label{coef-1-} \\
&=&\left(
\begin{array}{ccc}
b_{1} & b_{3} & b_{5} \\
b_{2} & b_{4} & b_{6}%
\end{array}%
\right) .  \label{coef-1}
\end{eqnarray}%
Clearly, Eq. (\ref{coef-1}) has proportional rows. Note that Eq. (\ref%
{coef-1}) is just Eq. (\ref{mt-2}).

Similarly, from Eqs. (\ref{cf-1-0}, \ref{rel-3}, \ref{order}), obtain the
corresponding coefficient matrix

\begin{eqnarray}
\left(
\begin{array}{ccc}
\alpha _{1}\beta _{1} & \alpha _{1}\beta _{2} & \alpha _{1}\beta _{3} \\
\alpha _{2}\beta _{1} & \alpha _{2}\beta _{2} & \alpha _{2}\beta _{3}%
\end{array}%
\right) &=&\left(
\begin{array}{ccc}
\mu _{1} & \mu _{2} & \mu _{3} \\
\mu _{4} & \mu _{5} & \mu _{6}%
\end{array}%
\right)  \label{coef-2-} \\
&=&\left(
\begin{array}{ccc}
b_{1} & b_{2} & b_{3} \\
b_{4} & b_{5} & b_{6}%
\end{array}%
\right) .  \label{coef-2}
\end{eqnarray}

Clearly, Eq. (\ref{coef-2}) has proportional rows. Note that Eq. (\ref%
{coef-2}) is just in Eq. (\ref{mt-3}).

From Eqs. (\ref{cf-1-0}, \ref{rel-5}, \ref{order}), obtain the corresponding
coefficient matrix
\begin{eqnarray}
\left(
\begin{array}{ccc}
\alpha _{1}\beta _{1} & \alpha _{1}\beta _{2} & \alpha _{1}\beta _{3} \\
\alpha _{2}\beta _{1} & \alpha _{2}\beta _{2} & \alpha _{2}\beta _{3}%
\end{array}%
\right) &=&\left(
\begin{array}{ccc}
\mu _{1} & \mu _{2} & \mu _{3} \\
\mu _{4} & \mu _{5} & \mu _{6}%
\end{array}%
\right)  \label{coef-3-} \\
&=&\left(
\begin{array}{ccc}
b_{1} & b_{3} & b_{4} \\
b_{2} & b_{5} & b_{6}%
\end{array}%
\right) .  \label{coef-3}
\end{eqnarray}

Clearly, Eq. (\ref{coef-3}) has proportional rows. Note that Eq. (\ref%
{coef-3}) is just in Eq. (\ref{mt-5}).

Example 1. For Case 1, let $|\Pi _{1}\rangle _{1...6}=(\frac{\sqrt{3}}{2}%
|000\rangle +\frac{1}{2}|111\rangle )_{346}\otimes (\frac{1}{\sqrt{3}}%
|000\rangle +\frac{1}{\sqrt{3}}|100\rangle +\frac{1}{\sqrt{3}}|111\rangle
)_{125},$

$|\Pi _{2}\rangle _{1...6}=(\frac{\sqrt{3}}{2}|000\rangle +\frac{1}{2}%
|111\rangle )_{123}\otimes (\frac{1}{\sqrt{3}}|000\rangle +\frac{1}{\sqrt{3}}%
|110\rangle +\frac{1}{\sqrt{3}}|111\rangle )_{456},$

$|\Pi _{3}\rangle _{1...6}=(\frac{\sqrt{3}}{2}|000\rangle +\frac{1}{2}%
|111\rangle )_{234}\otimes (\frac{1}{\sqrt{3}}|000\rangle +\frac{1}{\sqrt{3}}%
|110\rangle +\frac{1}{\sqrt{3}}|111\rangle )_{156}.$

Then, the corresponding coefficient matrices are the ones in Eqs. (\ref{mt-2}%
, \ref{mt-3}, \ref{mt-5}) which have proportional rows for $|\Pi _{i}\rangle
_{1...6}$, $i=1,2,3,$ respectively.

For Case 2, Eq. (\ref{form-2}) can be rewritten as
\begin{eqnarray}
|\psi \rangle _{1\cdots n} &=&\alpha _{1}\beta _{1}|\ell _{1}\rangle
_{1\cdots n}+\alpha _{1}\beta _{2}|\ell _{2}\rangle _{1\cdots n}+\alpha
_{1}\beta _{3}|\ell _{3}\rangle _{1\cdots n}+  \nonumber \\
&&\alpha _{2}\beta _{1}|\ell _{4}\rangle _{1\cdots n}+\alpha _{2}\beta
_{2}|\ell _{5}\rangle _{1\cdots n}+\alpha _{2}\beta _{3}|\ell _{6}\rangle
_{1\cdots n}  \label{cf-2-} \\
&=&\sum_{i=1}^{6}\mu _{i}|\ell _{i}\rangle _{1\cdots n},  \label{cf-2}
\end{eqnarray}%
where

\begin{eqnarray}
|\ell _{1}\rangle _{1\cdots n} &=&|0_{1}\gamma _{2}\cdots \gamma _{k}\rangle
_{p_{1}\cdots p_{k}}|0_{1}\sigma _{2}\cdots \sigma _{s}\rangle _{q_{1}\cdots
q_{s}}, \\
|\ell _{2}\rangle _{1\cdots n} &=&|0_{1}\gamma _{2}\cdots \gamma _{k}\rangle
_{p_{1}\cdots p_{k}}|0_{1}\tau _{2}\cdots \tau _{s}\rangle _{q_{1}\cdots
q_{s}}, \\
|\ell _{3}\rangle _{1\cdots n} &=&|0_{1}\gamma _{2}\cdots \gamma _{k}\rangle
_{p_{1}\cdots p_{k}}|1_{1}\upsilon _{2}\cdots \upsilon _{s}\rangle
)_{q_{1}\cdots q_{s}}, \\
|\ell _{4}\rangle _{1\cdots n} &=&|1_{1}\gamma _{2}^{\prime }\cdots \gamma
_{k}^{\prime }\rangle )_{p_{1}\cdots p_{k}}|0_{1}\sigma _{2}\cdots \sigma
_{s}\rangle _{q_{1}\cdots q_{s}}, \\
|\ell _{5}\rangle _{1\cdots n} &=&|1_{1}\gamma _{2}^{\prime }\cdots \gamma
_{k}^{\prime }\rangle )_{p_{1}\cdots p_{k}}|0_{1}\tau _{2}\cdots \tau
_{s}\rangle _{q_{1}\cdots q_{s}}, \\
|\ell _{6}\rangle _{1\cdots n} &=&|1_{1}\gamma _{2}^{\prime }\cdots \gamma
_{k}^{\prime }\rangle )_{p_{1}\cdots p_{k}}|1_{1}\upsilon _{2}\cdots
\upsilon _{s}\rangle )_{q_{1}\cdots q_{s}}.
\end{eqnarray}

It is easy to see that $\ell _{1}$ and $\ell _{4}$, $\ell _{2}$ and $\ell
_{5}$, and $\ell _{3}$ and $\ell _{6}$ are the desired three partially
complementary pairs. Note that $\sigma _{2}\cdots \sigma _{s}<\tau
_{2}\cdots \tau _{s}$. It is not hard to show that $\ell _{2}<\ell _{3}$, $%
\ell _{4}<\ell _{5}$, and $\ell _{2}<\ell _{5}$. \

\ We next show that it is not possible for $\ell _{4}<\ell _{2}$ and $\ell
_{3}<\ell _{5}$ to both hold as follows. When $p_{1}$ is qubit 1, $\ell
_{3}<\ell _{5}$ but $\ell _{4}>\ell _{2}$. When $q_{1}$ is qubit 1, $\ell
_{3}>\ell _{5}$. Therefore, it is not possible for $\ell _{4}<\ell _{2}$ and
$\ell _{3}<\ell _{5}$ to both hold. Specially, $\ell _{1}<\ell _{4}<\ell
_{2}<\ell _{3}<\ell _{5}<\ell _{6}$ does not hold.

Similarly, we show that $\ell _{4}<\ell _{3}<\ell _{5}$ does not hold. The
following is our argument. When $p_{1}$ is qubit 1, $\ell _{4}>\ell _{3}$.
When $q_{1}$ is qubit 1, $\ell _{5}<\ell _{3}$. \ Specially, that $\ell
_{1}<\ell _{2}<\ell _{4}<\ell _{3}<\ell _{5}<\ell _{6}$ does not hold.

Therefore, $\ell _{i},i=1,..,6,$ must satisfy one of the following

\begin{eqnarray}
\ell _{1} &<&\ell _{2}<\ell _{4}<\ell _{5}<\ell _{3}<\ell _{6}
\label{rela-1} \\
\ell _{1} &<&\ell _{4}<\ell _{2}<\ell _{5}<\ell _{3}<\ell _{6}
\label{rela-2} \\
\ell _{1} &<&\ell _{2}<\ell _{3}<\ell _{4}<\ell _{5}<\ell _{6}
\label{rela-3}
\end{eqnarray}

From Eqs. (\ref{cf-2}, \ref{rela-1}, \ref{order}), it is clear that $|\ell
_{1}\rangle =|B_{1}\rangle $, $|\ell _{2}\rangle =|B_{2}\rangle $, $|\ell
_{4}\rangle =|B_{3}\rangle $, $|\ell _{5}\rangle =|B_{4}\rangle $, $|\ell
_{3}\rangle =|B_{5}\rangle $, and $|\ell _{6}\rangle =$ $|B_{6}\rangle $.
Thus, obtain the following corresponding coefficient matrix
\begin{eqnarray}
\left(
\begin{array}{ccc}
\alpha _{1}\beta _{1} & \alpha _{1}\beta _{2} & \alpha _{1}\beta _{3} \\
\alpha _{2}\beta _{1} & \alpha _{2}\beta _{2} & \alpha _{2}\beta _{3}%
\end{array}%
\right) &=&\left(
\begin{array}{ccc}
\mu _{1} & \mu _{2} & \mu _{3} \\
\mu _{4} & \mu _{5} & \mu _{6}%
\end{array}%
\right)  \label{cm-2-} \\
&=&\left(
\begin{array}{ccc}
b_{1} & b_{2} & b_{5} \\
b_{3} & b_{4} & b_{6}%
\end{array}%
\right) .  \label{cm-2}
\end{eqnarray}

Clearly, Eq. (\ref{cm-2}) has proportional rows. Note that Eq. (\ref{cm-2})
is just Eq. (\ref{mt-1}).

Similarly, from Eqs. (\ref{cf-2}, \ref{rela-2}, \ref{order}), we obtain the
corresponding coefficient matrix

\begin{eqnarray}
\left(
\begin{array}{ccc}
\alpha _{1}\beta _{1} & \alpha _{1}\beta _{2} & \alpha _{1}\beta _{3} \\
\alpha _{2}\beta _{1} & \alpha _{2}\beta _{2} & \alpha _{2}\beta _{3}%
\end{array}%
\right) &=&\left(
\begin{array}{ccc}
\mu _{1} & \mu _{2} & \mu _{3} \\
\mu _{4} & \mu _{5} & \mu _{6}%
\end{array}%
\right)  \label{cm-3-} \\
&=&\left(
\begin{array}{ccc}
b_{1} & b_{3} & b_{5} \\
b_{2} & b_{4} & b_{6}%
\end{array}%
\right) .  \label{cm-3}
\end{eqnarray}

Clearly, Eq. (\ref{cm-3}) has proportional rows. note that Eq. (\ref{cm-3})\
is just in Eq. (\ref{mt-2}),

From Eqs. (\ref{cf-2}, \ref{rela-3}, \ref{order}), we obtain the
corresponding coefficient matrix

\begin{eqnarray}
\left(
\begin{array}{ccc}
\alpha _{1}\beta _{1} & \alpha _{1}\beta _{2} & \alpha _{1}\beta _{3} \\
\alpha _{2}\beta _{1} & \alpha _{2}\beta _{2} & \alpha _{2}\beta _{3}%
\end{array}%
\right) &=&\left(
\begin{array}{ccc}
\mu _{1} & \mu _{2} & \mu _{3} \\
\mu _{4} & \mu _{5} & \mu _{6}%
\end{array}%
\right)  \label{cm-4-} \\
&=&\left(
\begin{array}{ccc}
b_{1} & b_{2} & b_{3} \\
b_{4} & b_{5} & b_{6}%
\end{array}%
\right) .  \label{cm-4}
\end{eqnarray}

Clearly, Eq. (\ref{cm-4}) has proportional rows. Note that Eq. (\ref{cm-4})\
is just in Eq. (\ref{mt-3}). Q.E.D.

Example 2. For Case 2, let $|\Gamma _{1}\rangle _{1...6}=(\frac{\sqrt{3}}{2}%
|000\rangle +\frac{1}{2}|111\rangle )_{234}\otimes (\frac{1}{\sqrt{3}}%
|000\rangle +\frac{1}{\sqrt{3}}|010\rangle +\frac{1}{\sqrt{3}}|111\rangle
)_{156},$

$|\Gamma _{2}\rangle _{1...6}=(\frac{\sqrt{3}}{2}|000\rangle +\frac{1}{2}%
|111\rangle )_{456}\otimes (\frac{1}{\sqrt{3}}|000\rangle +\frac{1}{\sqrt{3}}%
|010\rangle +\frac{1}{\sqrt{3}}|111\rangle )123$,

$|\Gamma _{3}\rangle _{1...6}=(\frac{\sqrt{3}}{2}|000\rangle +\frac{1}{2}%
|111\rangle )_{123}\otimes (\frac{1}{\sqrt{3}}|000\rangle +\frac{1}{\sqrt{3}}%
|010\rangle +\frac{1}{\sqrt{3}}|111\rangle )456$.

Then, the corresponding coefficient matrices are the ones in Eqs. (\ref{mt-1}%
, \ref{mt-2}, \ref{mt-3}) which have proportional rows for $|\Gamma
_{i}\rangle _{1...6}$, $i=1,2,3,$ respectively.

Lemma 1 implies the following Corollaries 1 and 2.

Corollary 1. Assume that an $n$-qubit state $|\psi \rangle _{1\cdots n}$
with $m=6$\ is not a trivially separable state. If any one of Eqs. (\ref%
{mt-1}-\ref{mt-5}) does not have proportional rows, then $|\psi \rangle
_{1\cdots n}$ is genuinely\ entangled.

It is trivial to check if the matrices in Eqs. (\ref{mt-1}-\ref{mt-5}) have
proportional rows.

Example 3. Let us verify $|\Psi _{6}\rangle $ and $|\Xi _{6}\rangle $ are
genuinely entangled \cite{Osterloh}. For $|\Psi _{6}\rangle $, the matrices
in Eqs. (\ref{mt-1}-\ref{mt-5}) are

\begin{eqnarray}
\left(
\begin{array}{ccc}
b_{1} & b_{2} & b_{5} \\
b_{3} & b_{4} & b_{6}%
\end{array}%
\right) &=&\left(
\begin{array}{ccc}
b_{1} & b_{3} & b_{5} \\
b_{2} & b_{4} & b_{6}%
\end{array}%
\right) =\left(
\begin{array}{ccc}
b_{1} & b_{2} & b_{3} \\
b_{4} & b_{5} & b_{6}%
\end{array}%
\right)  \nonumber \\
&=&\left(
\begin{array}{ccc}
b_{1} & b_{3} & b_{4} \\
b_{2} & b_{5} & b_{6}%
\end{array}%
\right) =\frac{1}{2\sqrt{2}}\left(
\begin{array}{ccc}
1 & 1 & 1 \\
1 & 1 & \sqrt{3}%
\end{array}%
\right) .  \nonumber
\end{eqnarray}

One can see that the above matrices don't have proportional rows. By
Corollary 1, the state $|\Psi _{6}\rangle $ of five qubits is genuinely
entangled. Similarly, we can show that the state $|\Xi _{6}\rangle $ of six
qubits is genuinely entangled.

Corollary 2. Assume that an $n$-qubit state$\ |\psi \rangle _{1\cdots n}$
with $m=6$\ is not a trivially separable state. If basis states of $|\psi
\rangle _{1\cdots n}$ don't consist of the three partially complementary
pairs in Eqs. (\ref{par-1}-\ref{par-6}), then $|\psi \rangle _{1\cdots n}$
is genuinely entangled.

By Corollary 2, $|\Psi _{6}\rangle $ and $|\Xi _{6}\rangle $ in Example 3
are genuinely entangled.

\section{A sufficient condition for separability of states of $n$ qubits
with six non-zero coefficients}

Lemma 2. (sufficiency) \ Assume that an $n$-qubit state $|\psi \rangle
_{1\cdots n}$ with $m=6$\ is not a trivially separable state. For $|\psi
\rangle _{1\cdots n}$, if its basis states consist of the three partially
complementary pairs in Eqs. (\ref{par-1}-\ref{par-6}) and the corresponding
coefficient matrix has proportional rows, then $|\psi \rangle _{1\cdots n}$
is separable.

Proof. The state can be written as $|\psi \rangle _{1\cdots
n}=\sum_{i=1}^{6}b_{i}|B_{i}\rangle _{1\cdots n}$. In Eqs. (\ref{par-1}-\ref%
{par-6}), $\{\sigma _{i}$, $\tau _{i}$, $\upsilon _{i}\}=\{1,0\}$, $%
i=1,\cdots ,s$, because $|\psi \rangle _{1\cdots n}$ is not a trivially
separable state. Then, there are two cases. Case 1. $\tau _{1}=0$ and Case
2. $\tau _{1}=1$.

Case 1. Let $|\omega \rangle =|0_{1}\gamma _{2}\cdots \gamma _{k}\rangle
_{p_{1}\cdots p_{k}}$ and $|\omega ^{\prime }\rangle =|1_{1}\gamma
_{2}^{\prime }\cdots \gamma _{k}^{\prime }\rangle _{p_{1}\cdots p_{k}}$, $%
|\pi _{1}\rangle =|0_{1}\sigma _{2}\cdots \sigma _{s}\rangle _{q_{1}\cdots
q_{s}}$, $|\pi _{2}\rangle =|0_{1}\tau _{2}\cdots \tau _{s}\rangle
_{q_{1}\cdots q_{s}}$, and $|\pi _{3}\rangle =|1_{1}\upsilon _{2}\cdots
\upsilon _{s}\rangle _{q_{1}\cdots q_{s}}$. Assume that $\sigma _{2}\cdots
\sigma _{s}<\tau _{2}\cdots \tau _{s}$ without loss of generality.

Let
\begin{eqnarray}
|\chi \rangle _{1\cdots n} &=&(\alpha _{1}|\omega \rangle +\alpha
_{2}|\omega ^{\prime }\rangle )_{p_{1}\cdots p_{k}}\otimes (\beta _{1}|\pi
_{1}\rangle +\beta _{2}|\pi _{2}\rangle +\beta _{3}|\pi _{3}\rangle
)_{q_{1}\cdots q_{s}}  \nonumber \\
&=&\alpha _{1}\beta _{1}|\kappa _{1}\rangle _{1\cdots n}+\alpha _{1}\beta
_{2}|\kappa _{2}\rangle _{1\cdots n}+\alpha _{1}\beta _{3}|\kappa
_{3}\rangle _{1\cdots n}  \nonumber \\
&&+\alpha _{2}\beta _{1}|\kappa _{4}\rangle _{1\cdots n}+\alpha _{2}\beta
_{2}|\kappa _{5}\rangle _{1\cdots n}+\alpha _{2}\beta _{3}|\kappa
_{6}\rangle _{1\cdots n}  \nonumber \\
&=&\sum_{i=1}^{6}\mu _{i}|\kappa _{i}\rangle _{1\cdots n}.  \label{cf-3}
\end{eqnarray}

Then $|\chi \rangle _{1\cdots n}$\ is separable. If there are $\alpha _{1}$,
$\alpha _{2}$, $\beta _{1}$, $\beta _{2}$, and $\beta _{3}$ such that $|\chi
\rangle _{1\cdots n}=|\psi \rangle _{1\cdots n}$, then $|\psi \rangle
_{1\cdots n}$ is separable. Let $|\chi \rangle _{1\cdots n}=|\psi \rangle
_{1\cdots n}$. Then, $|\kappa _{1}\rangle =$ $|B_{1}\rangle $ and $|\kappa
_{6}\rangle =|B_{6}\rangle $. From Eq. (\ref{cf-3}), obtain
\begin{eqnarray}
&&\left(
\begin{array}{ccc}
\alpha _{1}\beta _{1} & \alpha _{1}\beta _{2} & \alpha _{1}\beta _{3} \\
\alpha _{2}\beta _{1} & \alpha _{2}\beta _{2} & \alpha _{2}\beta _{3}%
\end{array}%
\right) =\left(
\begin{array}{ccc}
\mu _{1} & \mu _{2} & \mu _{3} \\
\mu _{4} & \mu _{5} & \mu _{6}%
\end{array}%
\right)  \label{coe-1} \\
&=&\left(
\begin{array}{ccc}
b_{1} & \mu _{2} & \mu _{3} \\
\mu _{4} & \mu _{5} & b_{6}%
\end{array}%
\right) .  \label{coe-2}
\end{eqnarray}

One can see that $\mu _{i}\in \{b_{2},...,b_{5}\},$ $i=2,...,5$. Similar to
the discussion for Case 2 of Section 3, the corresponding coefficient matrix
in Eq. (\ref{coe-2}) is one of Eqs. (\ref{mt-1}, \ref{mt-2}, \ref{mt-3}). By
assumption, the corresponding coefficient matrix in Eq. (\ref{coe-2}) has
proportional rows. Via Theorem 1 in \ \cite{Li-24}, Eq. (\ref{coe-1}) has at
least one solution for $\alpha _{i}$ and $\beta _{j}$. Therefore, $|\psi
\rangle _{1\cdots n}$ is separable.

Case 2. Similarly, Lemma 2 holds. Q.E.D.

\section{A necessary and sufficient condition for genuinely\ entangled
states of $n$ qubits with six non-zero coefficients}

Lemmas 1 and 2 imply the following.

Theorem 1. Assume that an $n$-qubit state $|\psi \rangle _{1\cdots n}$ with $%
m=6$\ is not a trivially separable state. Then, $|\psi \rangle _{1\cdots n}$%
\ is separable if and only if its basis states consist of the three
partially complementary pairs in Eqs. (\ref{par-1}-\ref{par-6})\ and the
corresponding coefficient matrix has proportional rows.

The contrapositive of Theorem 1 leads to the following.

Theorem 2. Assume that an $n$-qubit state $|\psi \rangle _{1\cdots n}$ with $%
m=6$\ is not a trivially separable state. Then, $|\psi \rangle _{1\cdots n}$%
\ is genuinely entangled if and only if its basis states do not consist of
the three partially complementary pairs in Eqs. (\ref{par-1}-\ref{par-6})\
or though its basis states consist of the three partially complementary
pairs in Eqs. (\ref{par-1}-\ref{par-6}), the corresponding coefficient
matrix does not have proportional rows.

One can see that it is simple and intuitive to check if the corresponding
coefficient matrix has proportional rows and basis states consist of the
three partially complementary pairs.

Example 4. Let $|\Theta \rangle =\frac{1}{\sqrt{6}}(|0000\rangle
+|0010\rangle +|0101\rangle +|0111\rangle +|1010\rangle -|1111\rangle
)_{1234}$. The six basis states consist of the following partially
complementary pairs: $(00)_{24}(00)_{13},(11)_{24}(00)_{13}$; $%
(00)_{24}(01)_{13}$, $(11)_{24}(01)_{13}$; $(00)_{24}(11)_{13}$, $%
(11)_{24}(11)_{13}$. The corresponding coefficient matrix in Eq. (\ref{mt-1}%
) is

\begin{equation}
\left(
\begin{array}{ccc}
b_{1} & b_{2} & b_{5} \\
b_{3} & b_{4} & b_{6}%
\end{array}%
\right) =\frac{1}{\sqrt{6}}\left(
\begin{array}{ccc}
1 & 1 & 1 \\
1 & 1 & -1%
\end{array}%
\right) .  \label{ex-4}
\end{equation}

Clearly, the matrix in Eq. (\ref{ex-4}) does not have proportional rows.
Therefore, $|\Theta \rangle $ is genuinely entangled.

Theorem 2 implies the following.

Corollary 3. Here, $(\gamma _{1}\gamma _{2}\cdots \gamma _{n})$ and $(\gamma
_{1}^{\prime }\gamma _{2}^{\prime }\cdots \gamma _{n}^{\prime })$ are called
a complete complementary pair. Then, for an $n$-qubit state, if its basis
states consist of three complete complementary pairs, then the state is
genuinely\ entangled.

The proof of Corollary 3 is put in Appendix A.

But, if a state has four complete complementary pairs of basis states, then
the state may be separable. For example, the three-qubit state $\frac{1}{2%
\sqrt{2}}\sum_{i,j,k=0,1}|ijk\rangle =\frac{1}{2\sqrt{2}}(|000\rangle
+\cdots |111\rangle )$ has four complete complementary pairs of basis
states. One can see that the state is separable.

Example 5. For $G_{abcd}=\alpha (|0000\rangle +|1111\rangle )+\beta
(|0011\rangle +|1100\rangle )+\gamma (|0101\rangle +|1010\rangle )+\delta
(|0110\rangle +|1001\rangle )$ in \cite{Verstraete}, if just one of $\alpha
,\beta ,\gamma ,$ and $\delta $ vanishes, then $G_{abcd}$ is genuinely
entangled by Corollary 3. The result does not appear in \cite{Verstraete}.

\section{Discussion}

Let $|\psi \rangle _{1\cdots n}$ be a pure state of $n$ qubits\ and $m$\ be
the number of non-zero coefficients of $|\psi \rangle _{1\cdots n}$, where $%
m=2p$, where $p$ is a prime. We discuss a necessary and sufficient condition
for the genuinely entangled states below.

The condition needs the following $p$ partially complementary pairs.

\begin{eqnarray*}
&&(0_{1}\gamma _{2}\cdots \gamma _{k})_{p_{1}\cdots p_{k}}(\sigma
_{1}^{(1)}\sigma _{2}^{(1)}\cdots \sigma _{s}^{(1)})_{q_{1}\cdots q_{s}}, \\
&&(1_{1}\gamma _{2}^{\prime }\cdots \gamma _{k}^{\prime })_{p_{1}\cdots
p_{k}}(\sigma _{1}^{(1)}\sigma _{2}^{(1)}\cdots \sigma
_{s}^{(1)})_{q_{1}\cdots q_{s}}; \\
&&\cdots \\
&&(0_{1}\gamma _{2}\cdots \gamma _{k})_{p_{1}\cdots p_{k}}(\sigma
_{1}^{(p)}\sigma _{2}^{(p)}\cdots \sigma _{s}^{(p)})_{q_{1}\cdots q_{s}} \\
&&(1_{1}\gamma _{2}^{\prime }\cdots \gamma _{k}^{\prime })_{p_{1}\cdots
p_{k}}(\sigma _{1}^{(p)}\sigma _{2}^{(p)}\cdots \sigma
_{s}^{(p)})_{q_{1}\cdots q_{s}},
\end{eqnarray*}%
where $k+s=n$, $\gamma _{i}$ and $\sigma _{k}^{(j)}$ are 0 or 1. Note that $%
\{\sigma _{k}^{(1)},\sigma _{k}^{(2)},\cdots ,\sigma _{k}^{(p)})=\{0,1\}$, $%
k=1,\cdots ,s$, because $|\psi \rangle _{1\cdots n}$ is not a trivially
separable state.

Assume that $|\psi \rangle _{1\cdots n}$ is not trivially separable. Then if
$|\psi \rangle _{1\cdots n}$ is separable, then it can be written as
\begin{eqnarray}
|\psi \rangle _{1\cdots n} &=&(\alpha _{1}|0_{1}\gamma _{2}\cdots \gamma
_{k}\rangle +\alpha _{2}|1_{1}\gamma _{2}^{\prime }\cdots \gamma
_{k}^{\prime }\rangle )_{p_{1}\cdots p_{k}}\otimes  \nonumber \\
&&(\beta _{1}|\sigma _{1}^{(1)}\sigma _{2}^{(1)}\cdots \sigma
_{s}^{(1)}\rangle +\cdots +\beta _{p}|\sigma _{1}^{(p)}\sigma
_{2}^{(p)}\cdots \sigma _{s}^{(p)}\rangle )_{q_{1}\cdots q_{s}}.
\label{dis-1}
\end{eqnarray}

We can rewrite $|\psi \rangle _{1\cdots n}$ in Eq. (\ref{dis-1}) as follows.

\begin{eqnarray*}
|\psi \rangle _{1\cdots n} &=&\alpha _{1}\beta _{1}|l_{1}\rangle _{1\cdots
n}+\alpha _{1}\beta _{2}|l_{2}\rangle _{1\cdots n}+\cdots +\alpha _{1}\beta
_{p}|l_{p}\rangle _{1\cdots n}+ \\
&&\alpha _{2}\beta _{1}|l_{p+1}\rangle _{1\cdots n}+\alpha _{2}\beta
_{2}|l_{p+2}\rangle _{1\cdots n}+\cdots +\alpha _{2}\beta _{p}|l_{2p}\rangle
_{1\cdots n},
\end{eqnarray*}%
where
\begin{eqnarray*}
|l_{1}\rangle _{1\cdots n} &=&|0_{1}\gamma _{2}\cdots \gamma _{k}\rangle
_{p_{1}\cdots p_{k}}|\sigma _{1}^{(1)}\sigma _{2}^{(1)}\cdots \sigma
_{s}^{(1)}\rangle _{q_{1}\cdots q_{s}}, \\
&&\cdots \\
|l_{p}\rangle _{1\cdots n} &=&|0_{1}\gamma _{2}\cdots \gamma _{k}\rangle
_{p_{1}\cdots p_{k}}|\sigma _{1}^{(p)}\sigma _{2}^{(p)}\cdots \sigma
_{s}^{(p)}\rangle _{q_{1}\cdots q_{s}}, \\
|l_{p+1}\rangle _{1\cdots n} &=&|1_{1}\gamma _{2}^{\prime }\cdots \gamma
_{k}^{\prime }\rangle )_{p_{1}\cdots p_{k}}|\sigma _{1}^{(1)}\sigma
_{2}^{(1)}\cdots \sigma _{s}^{(1)}\rangle _{q_{1}\cdots q_{s}}, \\
&&\cdots \\
|l_{2p}\rangle _{1\cdots n} &=&|1_{1}\gamma _{2}^{\prime }\cdots \gamma
_{k}^{\prime }\rangle )_{p_{1}\cdots p_{k}}|\sigma _{1}^{(p)}\sigma
_{2}^{(p)}\cdots \sigma _{s}^{(p)}\rangle _{q_{1}\cdots q_{s}}.
\end{eqnarray*}

Thus, obtain the following corresponding coefficient matrix
\[
\left(
\begin{array}{ccc}
\alpha _{1}\beta _{1} & \cdots & \alpha _{1}\beta _{p} \\
\alpha _{2}\beta _{1} & \cdots & \alpha _{2}\beta _{p}%
\end{array}%
\right) =\left(
\begin{array}{ccc}
\varsigma _{1} & \cdots & \varsigma _{p} \\
\varsigma _{p+1} & \cdots & \varsigma _{2p}%
\end{array}%
\right) .
\]

What to do next is how to find the forms of the corresponding coefficient
matrices. When $m=4$, there are two forms \cite{Li-24}. When $m=6$, there
are four forms. How many forms of the corresponding coefficient matrices are
there when $m=2p$?

\section{Summary}

In this paper, we propose a necessary and sufficient condition for the
separability of pure states of $n$ qubits with six non-zero coefficients.
The contrapositive of this result reads a necessary and sufficient condition
for genuinely entangled states of $n$ qubits with six non-zero coefficients.
The conditions are simple and intuitive. It is easy to use the conditions to
verify that Osterloh and Siewert's state $|\Psi _{6}\rangle $ of five qubits
and the state $|\Xi _{6}\rangle $ of six qubits are genuinely entangled.

\section{Appendix A The proof of Corollary 3}

Proof. Assume that $\gamma _{1}\gamma _{2}\cdots \gamma _{n}$ and $\gamma
_{1}^{\prime }\gamma _{2}^{\prime }\cdots \gamma _{n}^{\prime }$, $\tau
_{1}\tau _{2}\cdots \tau _{n}$ and $\tau _{1}^{\prime }\tau _{2}^{\prime
}\cdots \tau _{n}^{\prime }$, and $\sigma _{1}\sigma _{2}\cdots \sigma _{n}$
and $\sigma _{1}^{\prime }\sigma _{2}^{\prime }\cdots \sigma _{n}^{\prime }$
are three complete complementary pairs. Assume that $\gamma _{1}\gamma
_{2}\cdots \gamma _{n}$ and $\tau _{1}\tau _{2}\cdots \tau _{n}$ are
partially complementary. For example, $\tau _{1}\tau _{2}\cdots \tau
_{n}=\gamma _{1}^{\prime }\cdots \gamma _{k}^{\prime }\gamma _{k+1}\cdots
\gamma _{n}$, where $\tau _{i}=\gamma _{i}^{\prime }$, $i=1,\cdots k$, and $%
\tau _{j}=\gamma _{j}$, $j=k+1,\cdots ,n$. Then, $\tau _{1}^{\prime }\tau
_{2}^{\prime }\cdots \tau _{n}^{\prime }=\gamma _{1}\cdots \gamma _{k}\gamma
_{k+1}^{\prime }\cdots \gamma _{n}^{\prime }$. Clearly, $\gamma _{1}^{\prime
}\gamma _{2}^{\prime }\cdots \gamma _{n}^{\prime }$ and $\tau _{1}^{\prime
}\tau _{2}^{\prime }\cdots \tau _{n}^{\prime }$ are partially complementary.
But, $\sigma _{1}\sigma _{2}\cdots \sigma _{n}$ and $\sigma _{1}^{\prime
}\sigma _{2}^{\prime }\cdots \sigma _{n}^{\prime }$ are not partially
complementary.

Statements and declarations: No financial interests, no competing interests,
no financial supports.

A data availability statement: It includes all data in the main text.


\begin{thebibliography}{99}
\bibitem{Li-24} D. Li, n qubits can be entangled in two different ways,
e-print, arxiv. 2406.19119.
\end{thebibliography}
\end{document}